# Noncentrosymmetric structural transitions in ultrashort ferroelectric AGaO$_3$/A′GaO$_3$ superlattices

Nayoung Song[1], James M. Rondinelli[2], and Bog G. Kim[1*]

[1]Department of Physics, Pusan National University, Pusan, 609-735, South Korea

[2] Department of Material Science and Engineering, Northwestern University, 2220 Campus Drive, Evanston, IL 60208-3108, USA

The effect of octahedral tilting on the acentric structural transitions in AGaO$_3$/A′GaO$_3$ [001], [110], and [111] superlattices (A, A' = La, Pr, Nd) is studied using density functional theory. We find the displacive transitions are driven by two octahedral rotations modes ($a^-a^-c^0$ and $a^0a^0c^+$ tilting), with amplitudes that depend on the A and A′ chemistry and cation ordering direction. We find the ground states structures of the [001] and [111] ordered superlattices are polar. The coupling of octahedral tilting modes through a hybrid improper ferroelectric mechanism induces the polar displacements and produces the macroscopic electric polarizations.

The tilting of transition metal oxygen octahedra can have profound effects on the physical properties of perovskite oxides, because the tilting transition can anharmonically couple to hard infrared modes to stabilize previously unknown low-symmetry acentric phases [1-3]. Among such available interactions, hybrid improper ferroelectricity (HIF) is of current interest in ultrashort period perovskite oxide superlattices as it is proven theoretical route to stabilize polar structures by interleaving nominally non-polar perovskites that exhibit phases with octahedral rotations [4-6]. With the recent developments in thin film oxide synthesis with layer-by-layer control [7, 8], new recommendations for the chemical composition and cation ordering direction in hybrid improper ferroelectric superlattices is needed. Theoretical understanding of various phase transition in the hybrid improper ferroelectricity is of crucial importance for understanding the nature of competing and cooperating polar and non-polar structures and designing electric-field switchable materials [9, 10].

Three dimensional networks of octahedral tilting in perovskite oxides ($ABO_3$) may be classified using Glazer notation [1, 3, 11-18]. The octahedral tilting uses the syntax, $a^{\#}b^{\#}c^{\#}$, in which the literals refer to tilts about the [100], [010], and [001] directions of the cubic perovskite (space group $Pm\bar{3}m$) and the amplitude of tilting. The superscript # is used to indicate if no (0) tilt or tilts of successive octahedra in the same (+) or opposite (–) sense occur in the structure. The $a^0a^0c^+$ tilting results in a perovskite with tetragonal symmetry, whereas the orthorhombic phase adopted by many perovskites exhibits the $a^-a^-c^+$ tilting pattern. The observed tilting pattern in oxide perovskites originates from the delicate balance of interatomic forces and ionic-covalent chemical bonding. The change of interatomic forces by atomic substitution results in the change of the tilting patterns, and thus the electronic properties of oxide materials owing to electron-lattice coupling.

Octahedral tilting plays an important role in the stability of ferroelectric phases for various perovskite derived systems [4, 5, 9, 10, 19-23]. Bousquet *et al.* [4] introduced the concept of ferroelectricity produced from anharmonic lattice interactions in an artificial superlattice of ferroelectric (PbTiO$_3$) and paraelectric (SrTiO$_3$) materials. Shortly afterwards, Benedek and Fennie illustrated that this hybrid improper ferroelectricity could exist in (ABO$_3$)$_2$(AO) layered perovskites [20]. Many recent works also focused on octahedral rotation induced ferroelectricity in cation ordered perovskites, (ABO$_3$)/(A'BO$_3$) systems, which also relies on a hybrid improper coupling mechanism [9, 10, 19-21]. Mulder et al. [9] introduced a design concept for large spontaneous polarizations in such oxides, showing that the spontaneous polarization is proportional established perovskite crystallographic descriptors: the tolerance difference multiplied by one minus average tolerance factor of parent compound. More recently, one of us identified the required rotational patterns conducive to "geometry" ferroelectricity in (A, A')B$_2$O$_6$ perovskite oxides with A cation order along [001], [110], and [111] directions [21]. Note that the [001] ordering direction is equivalent to an ultrashort period superlattice, (ABO$_3$)/(A'BO$_3$), grown along the [001] direction.

Here, we report on the ground state structures of bulk perovskite AGaO$_3$ compounds and AGaO$_3$/A'GaO$_3$ superlattices (A, A' = La, Pr, Nd) with [001], [110], and [111] cation ordering obtained from density functional theory simulations. As indicated from prior studies on AGaO$_3$ [11, 24, 25], these lanthanide gallates exhibit orthorhombic phases, where the three dimensional network of the octahedral tilting is characterized by the $a^-a^-c^+$ tilt pattern in Glazer notation, i.e., in-phase rotations along the long crystallographic axis and out-of-phase rotations about the two other orthogonal directions. After optimizing the structure of the nine configurations, we applied symmetry mode analyses to understand the role of the GaO$_6$ tilt modes in stabilizing the low-symmetry structures. For three combinations of A site atoms (La/Pr, La/Nd, and Pr/Nd) and in two ordered structures ([001] and [111]), we found polar ground states and computed the electric polarizations. Analysis of the ionic displacements in these systems reveals that the sizeable polarizations predominately arise from A and A' cation displacements. This finding is consistent with the theory of hybrid improper ferroelectricity being operative in ultra-short period superlattices of orthorhombic perovskites and provides an additional family of artificial perovskite oxides for experimentation.

We performed first principles density functional calculations with the PBEsol generalized gradient approximation (GGA) [26] and the projector-augmented-wave method as implemented in VASP [27, 28]. We considered the following valence electron configuration: $5s^26s^25p^65d^1$ for La, $5s^26s^25p^65d^1$ for Pr, $5s^26s^25p^65d^1$ for Nd, $3d^{10}4s^24p^1$ for Ga, and $2s^22p^4$ for Oxygen. The electronic wave functions are expanded with plane waves up to a kinetic-energy cutoff of 400 eV except for structural optimization, where a kinetic energy cutoff of 520 eV has been applied in order to reduce the effect of Pulay stress. Momentum space integrations are performed using a $4 \times 5 \times 5$ Γ-centered Monkhorst-Pack $k$-mesh [29]. For the various symmetries examined, the lattice constants and internal coordinates were fully optimized until the residual Hellmann-Feyman forces became smaller that $10^{-1}$ meV/Å. The phonon dispersion curves and phonon partial density of state were obtained using the frozen phonon method and the PHONOPY program for pre- and post- processing [30]. The spontaneous electric polarization was obtained by using the Berry phase method [31]. Note we also computed the spontaneous polarization using the born effective charges obtained from density functional perturbation theory [32]. The ISOTROPY and AMPLIMODES program were utilized to verify the group-subgroup relationships and perform the structural mode-decompositions [33, 34].

In order to investigate possible low-symmetry structures for the $AGaO_3/A'GaO_3$ superlattices, we extended the basic 5-atom cubic unit cell to $2 \times 2 \times 2$ supercell containing 40 atoms [Fig. 1(a)]. Simple cation ordering on the A-site can be achieved by in three variants: a [001] layered configuration, a [110] columnar configuration, and finally a [111] 'rocksalt' configuration [21] as shown in Fig. 1(b), 1(c), and 1(d), respectively. With three different configurations of A-site ordering and three different A-site atom pairing combinations (La/Pr, La/Nd, and Pr/Nd), we have optimized 9 ordered perovskite variants in both a high symmetry phase, containing neither octahedral rotations nor distortions, and low-symmetry phases, exhibiting such modes, in our calculation. Note that in our simulations the non-primitive cell depicted in Figure 1 are replaced with primitive cells containing 20 atoms and 10 atoms depending on ordering sequences for computational efficiency.

The results of our DFT calculations are summarized in Fig. 2, whereby the phase transition sequence of the superlattices is given based on crystallographic symmetry and relevant order parameters across the structural transitions. In the [001] layered configuration, the bi-colored ordering of the A-site lattice produces a tetragonal space group, $P4/mmm$, for $AGaO_3/A'GaO_3$. Group theory analysis shows that $a^-a^-c^+$ octahedral tilting in the [001] superlattice leads to a further symmetry reduction to the polar $Pmc2_1$ orthorhombic phase (see Fig. 2) [9, 10]. All possible superlattice combinations for the parent compounds in [001] layered configurations (La/Pr, La/Nd, and Pr/Nd) have been examined, and the calculated equilibrium structural parameters and electric polarizations are summarized in the Table 1. We find the polar orthorhombic phase ($Pmc2_1$) is always more stable than the tetragonal phase ($P4/mmm$) for all three superlattices with the [001] layered configuration, which is a result of the orthorhombic instability (tilting instability) of three compounds. $\Delta E_{hl}$ is the energy difference between the high-symmetry phase and the low-symmetry phase and the stability of the polar $Pmc2_1$ increases as the tolerance factor of A-site atoms deviates from unity, supporting the concept that the orthorhombic instability is connected to the tolerance factor. Figure 2 also summarizes the anticipated phase transitions sequences for the [110] and [111] superlattices. The [111] superlattice without octahedral tilting belongs to a centrosymmetric $Fm\overline{3}m$ symmetry; the $a^-a^-c^+$ tilting lowers the symmetry to an orthorhombic $Pmn2_1$ phase. For the [110] ordered superlattice, the high symmetry phase is orthorhombic $Pmmm$, and whereas unlike the two previously described superlattices, the low symmetry phase is *centrosymmetric* ($P2_1/m$).

The tilting angles for the low symmetry phases are calculated and tabulated in Table 1. It is smallest for La/Pr superlattice and largest for Pr/Nd superlattice. The magnitude of the tilting angle for the superlattices can be understood from the tendency of the parent gallates to undergo octahedral rotations as described by the crystallographic tolerance factor (See Suppl. Fig. S1 and Table S1 for the detailed information on parent compounds). We computed the tolerance factor for the parent compound using both Shannon ionic radii [35], which give 0.966, 0.950, and 0.935, and that using bond valence parameters, 0.956, 0.944, and 0.932, for $LaGaO_3$, $PrGaO_3$, and $NdGaO_3$, respectively [36], Consistent with the deviation from unity, $NdGaO_3$ exhibits the largest octahedral rotations (Suppl. Fig. S1f). The tolerance factor for each superlattice, calculated from the average of two bulk parent compounds, also correlates well with the tilting angles of the superlattices (See Suppl. Fig. S2 for the detailed calculation of [001] ordered superlattice and Table S2 for the atomic Wyckoff position of superlattices). We find that the superlattice tilt angles, obtained by averaging the Ga-O-Ga angles of each bulk gallate, are slightly overestimated. This occurs because the equilibrium tilting angles are sensitive to the unit cell volume of the perovskite structure [37], and the average unit cell volumes of the parent compounds are underestimated compared to the fully relaxed superlattice. The energy difference between the orthorhombic phase and the tetragonal phase is also closely related with the octahedral tilt instability. $\Delta E_{hl}$ is the energy difference between the high symmetry phase and the low symmetry phase. Indeed, $\Delta E_{hl}$ increases as the tilting angle increases from La/Pr to Pr/Nd [9, 10, 38].

We now examine the displacive modes responsible for the symmetry reduction using a crystallographic mode based analysis and representation theory. We have utilized the ISOTROPY program as well as AMPLIMODE program to facilitate the analysis of the transition pathways [33, 34]. The amplitudes of the modes active across the transition are summarized in Table 2 and correspond to the labels presented in Figure 2. All symmetry labels are defined relative to the direct supergroup of the phase as specified in Fig. 2. For the [001] ordered structure, the high symmetry phase is *P4/mmm* and low symmetry phase is *Pmc*$2_1$ and the $M_3^+$, $M_5^-$ and $\Gamma_5^-$ are the relevant modes across the phase transition (See Suppl. Fig. S3 for the atomic movement in each mode of the [001] ordered superlattice). The $M_3^+$ and $M_5^-$ zone boundary modes are the cell doubling distortions (octahedral tilting) and are the primary order parameters for the structural transition since they make the greatest contributions to energetically stabilizing the polar ground state (see discussion below). The $\Gamma_5^-$ mode is an additional mode permitted in the [001] superlattice because inversion symmetry is already broken by the two A-site atoms. In the structural phase transition, the non-zero and small amplitude of $\Gamma_5^-$ mode indicates that it is likely a secondary order parameter and HIF is active in the superlattice [5, 9, 10]. We confirm this through total energy calculations as a function of mode amplitude described below. For the [110] ordered superlattice, the high symmetry phase is orthorhombic *Pmmm* phase and the low-symmetry phase is monoclinic P$2_1$/m phase, where the $Z_5^+$ and $\Gamma_4^+$ are the relevant modes involved in the phase transition. The [111] superlattice adopts the orthorhombic *Pmn*$2_1$ phase with modes transforming as $\Gamma_5^-$, $\Gamma_4^-$, $X_2^+$, and $X_5^-$ modes contributing to the symmetry reduction from the cubic F$m\overline{3}m$ phase without any tilting. (See Suppl. Fig. S4 for the atomic movement in each mode of the [111] superlattice). It is interesting to note that the *Pmc*$2_1$ ([001] superlattice) and the P*mn2$_1$* ([111] superlattice) are polar phases with spontaneous electric polarizations (indicated with yellow boxes).

To explore if hybrid ferroelectricity is active in the polar polymorphs, we plot in Figure 3 the energy change with respect to the amplitude of each symmetry mode for the [001] and [111] La/Nd superlattices. In each superlattice, zero mode amplitude corresponds to an idealized superlattice with lattice constants identical to the ground state structure but without any octahedral distortions or atomic displacements. For the [001] ordering, the relevant symmetry modes are the $M_3^+$, $M_5^-$ and $\Gamma_5^-$ modes (Fig. 2 and Fig. 3(a)). The total energy of the system exhibits a double-well potential energy surface for the two octahedral tilting modes, which indicates spontaneous symmetry breaking towards a structure with octahedral rotations is favored. The energy related with the $M_5^-$ tilting mode is about 0.685 eV/f.u. (1 formula unit for the superlattice corresponds to a cell with 10 atoms) and that of the $M_3^+$ tilting mode is about 0.292 eV/f.u. The tilting energies associated with the $M_5^-$ and $M_3^+$ modes are significant, indicating that a transition to a paraelectric tetragonal phase is unlikely to occur under ambient condition until high temperature. Compared to the $M_5^-$ and $M_3^+$ tilting modes, the energy related with the $\Gamma_5^-$ mode is very small (it is nearly zero within our calculation accuracy), indicating the mode is dynamically stable. The phase transition to the ground state $Pmc2_1$ phase is achieved by the coupling among these three modes (indicated as all in Fig. 3(a)) and the energy associated with the transition is 0.769 eV/f.u. Interestingly, this value is less than that obtained by the direct sum of the individual modes, which indicates that while all three modes coexist and are coupled, there is some competition among the $M_5^-$, $M_3^+$ and $\Gamma_5^-$ modes which have amplitudes of 1.2704, 0.7982, and 0.4780 Å, respectively.

For the [111] ordering, the $\Gamma_5^-$, $\Gamma_4^-$, $X_2^+$, and $X_5^-$ modes are obtained from our mode decomposition analysis (Fig. 2 and Fig. 3 (b)). The total energy as a function of mode amplitude is also described by double-well type of potential. Here, the largest contribution comes from $\Gamma_5^-$ and $X_2^+$ modes which correspond to the $GaO_6$ tilt transition. In this phase transition, the modes are described by different symmetry labels but have the same physical meaning as those described in the [001] superlattice: Our analysis shows that $\Gamma_5^-$ is related with the $c^+$ tilting and $X_2^+$ is connected to the $a^-$ tilt mode. The mode associated with ferroelectric polarization is $\Gamma_4^-$ mode. It is interesting to note that the total amplitude is nearly same for [001] and [111] superlattice, however, the mode amplitude related to the ferroelectric transition ($\Gamma_5^-$ for the [001] superlattice and $\Gamma_4^-$ for the [111] superlattice) is approximately 25% larger for the [001] A and A' cation arrangement (see Table 2).

We now discuss the phase stability and phase transition of the superlattices. In order to check the dynamic stability of the high symmetry phases and the low-symmetry phases, phonon dispersion curves are calculated for each ground state superlattice variant. The calculations of the phonon dispersion curves were performed using the force constant method [30] in a 2 × 2 × 2 supercell containing 80 or 160 atoms for the high symmetry and low symmetry superlattice geometries, respectively. The force constants were calculated for the displacement of atoms of up to 0.04 Å and the dynamical matrix at each *q* point in the Brillouin zone was constructed by Fourier transforming the force constant matrix calculated at the Γ point and the zone boundaries. The high symmetry phases of the [001], [110], and [111] configurations have the imaginary unstable phonon modes at zone boundary, which is associated with the energetically favorable octahedral tilt distortions. Fig. 3(c) and 3(d) depict the phonon dispersion curves for the equilibrium [001] and [111] configurations of the La/Nd superlattices in the polar structures (see Suppl. Fig. S5 for the phonon dispersion curves of La/Pr and Pr/Nd). We find that all dispersion curves have real mode frequencies and there are no unstable modes in the low symmetry phase, indicating dynamical stability of the phases.

Fig. 4(a) shows the spontaneous polarization for three [001] and [111] superlattices. We have used a linearized form of the total polarization of a crystal that combines Born effective charges and displacements of ions from ideal positions [32] and find that the polarization values obtained from this method and the Berry phase method [31, 39] are nearly identical. We find that in the systems surveyed the [111] superlattices consistently have a lower polarization than the [001] superlattices. To understand the microscopic origin of this behavior, we examine the spontaneous polarization of the lower symmetry phases for each composition in the [001] and [111] superlattices in a layer-decomposed manner using the Born effective charges for each ion obtained from the equilibrium polar structures [32]. Figure 4 depicts the spontaneous polarization, $P_s$, results for [001] and [111] ordering. For [001] ordering in the La/Nd superlattice (Fig. 4(b) and 4(c)), the polarizations from the NdO layer and LaO layers are 13.8μC/cm$^2$ and -11.3 μC/cm$^2$, which gives a net polarization of 2.60 μC/cm$^2$. The polarization from each GaO$_2$ layer is much smaller, 1.38 μC/cm$^2$, and is found to always be aligned in the direction of the NdO layer. The total polarization for the [001] superlattice is ~ 5.45 μC/cm$^2$ with approximately 50% of that value owing to contributions from the LaO/NdO layers and the remainder from the two GaO$_2$ layers. As the tolerance difference between the parent perovskites decreases from La/Nd, La/Pr to Pr/Nd, the spontaneous polarization also decreases from 5.45 4.03, to 1.50 μC/cm$^2$, respectively.

Now we examine the spontaneous polarization for the [111] ordered variants (Fig. 4(d) and 4(e)). For the [111] superlattices, four constituent layers are needed to describe the layer polarization structure: two LaO/NdO and two $GaO_2$ layers. As indicated by the arrows in Fig. 4(dc), each AO/A'O layer contains La and one Nd cations. The polarization from the NdO layer is 9.34 $\mu C/cm^2$ and the polarization from the LaO layer is -9.24 $\mu C/cm^2$ and their contribution to the total polarization therefore nearly cancels, i.e., the net polarization from the A-cation sublattice, AO/A'O, is 0.104 $\mu C/cm^2$ in each plane. The polarization from the $GaO_2$ layer is 1.21 $\mu C/cm^2$, which gives a net polarization for the [111] superlattices of 2.63 $\mu C/cm^2$. Unlike the [001] superlattices, 92% of the total polarization derives from the $GaO_2$ layer. In the first approximation, for the [001] superlattices, the net polarization has equal contributions from the AO layer and $BO_2$ layer and, for [111] superlattices, the net polarization has dominant contribution from $BO_2$ layer and very small contribution from AO layer.

In conclusion, we have presented detailed analyses of the structural phases of $AGaO_3$/$A'GaO_3$ [001], [110], and [111] superlattices and described the microscopic origins of the polar phases in the [001] and [111] variants using first-principles electronic structure calculations. We find the tendency of the cation ordered structures to undergo displacive phase transitions is correlated with the susceptibility of the bulk parent compounds to exhibit the same rotational distortions, which we explain using crystal-chemistry concepts. The structural phase transitions of each superlattice are analyzed by group theory and analysis of the polar phase transitions is accomplished with a symmetry-mode analysis. Exploration of the potential energy surface reveals that the spontaneous inversion symmetry breaking from the high symmetry phase to low symmetry phase is largely driven by the rotations of $GaO_6$ octahedra. We also computed the spontaneous electric polarization for [001] and [111] superlattices with space group of *Pmc*$2_1$ ([001] superlattice) and the *Pmn*$2_1$ ([111] superlattice). Our layer decomposed polarization analysis explains the observation that the polarization values for the [001] superlattices is larger than that of the [111] superlattice in $AGaO_3$/$A'GaO_3$

This study was supported by NSF of Korea (NSF-2013R1A1A2004496). The computational resources have been provided by KISTI Supercomputing Center (Project No. KSC-2013-C1-029). JMR acknowledges support from the Penn State Center for Nanoscience under grant no. NSF-DMR-0820404.

# Figure Captions

Figure 1. The structure of perovskite with three types of cation ordering, (a) Simple cubic phase of $AGaO_3$. $AGaO_3/A'GaO_3$ (A,A'=La,Pr,Nd) superlattice with (b) [001] layered ordering, (c) [110] columnar ordering, and (d) [111] rocksalt ordering.

Figure 2. High and low symmetry phases of [001], [110], and [111] $AGaO_3/A'GaO_3$ (A,A'=La, Pr, Nd) superlattices. The symmetry labels of the mode appearing across the phase transition are also indicated and the shaded (yellow) boxed phases are polar structures compatible with spontaneous electric polarizations. All symmetry labels are defined relative to the phase immediately above the label.

Figure 3. The energy surface as a function of mode amplitude for the (a) [001] ordered and (b) [111] ordered $LaGaO_3/NdGaO_3$ superlattices. Phonon dispersion curve in the low symmetry phase of (c) [001] ordered and (d) [111] ordered $LaGaO_3/NdGaO_3$ superlattices.

Figure 4. (a) Spontaneous electric polarizations for the [001] and [111] La/Nd. Pr/Nd, and La/Pr superlattices. Note that the spontaneous polarization of the [111] superlattices show nearly half the total value of that of found for the [001] superlattices because of the partial cancellation in the layer dipoles appear in the in AO/A'O monoxide planes. Cation displacement patterns relevant to spontaneous polarization of (b) [001] and (d) [111] superlattices shown in the *a-b* plane. Layer decomposed spontaneous polarization of (c) [001] and (e) [111] superlattices.

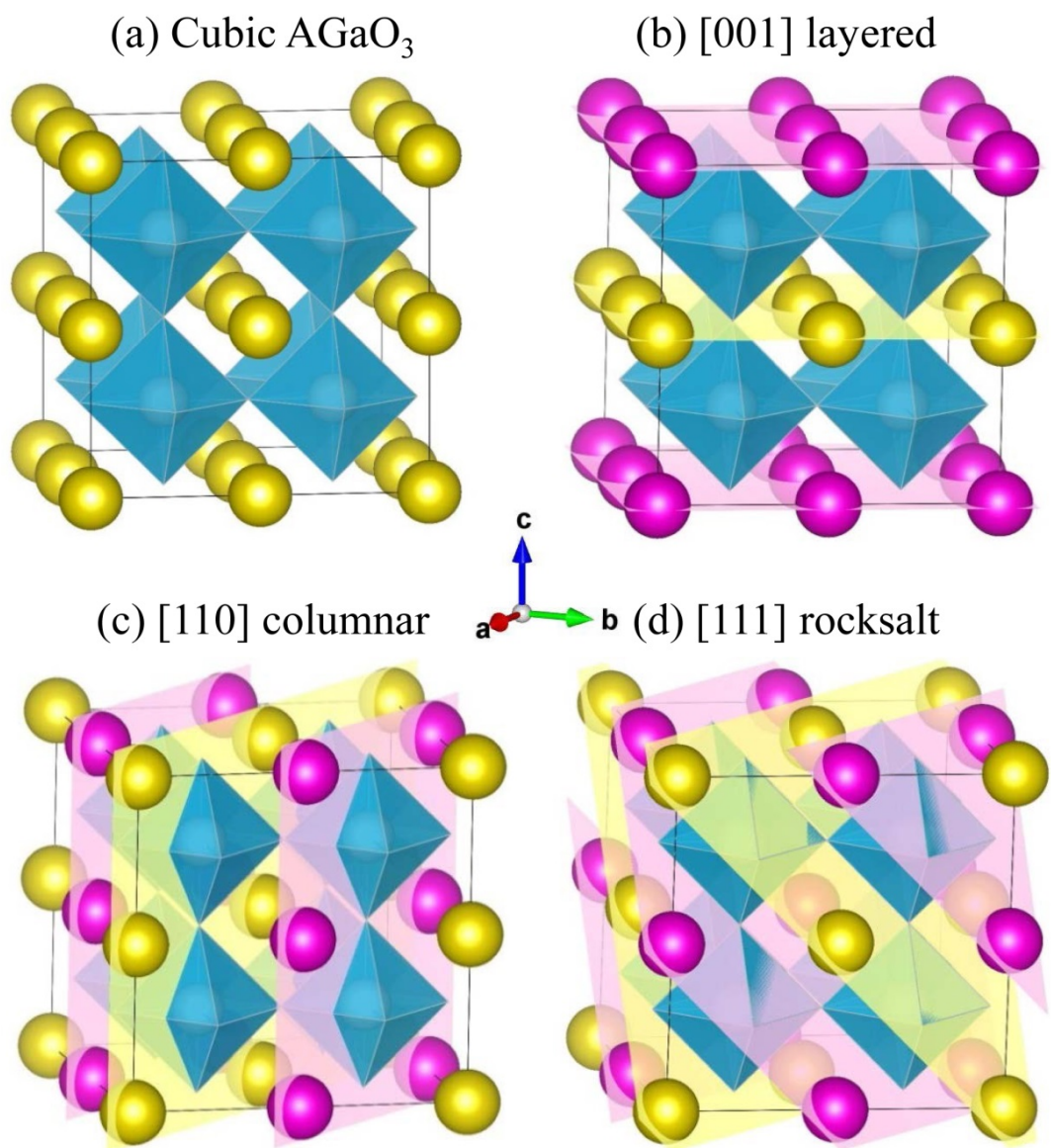

Figure 1 (Color online) Song *et al.*

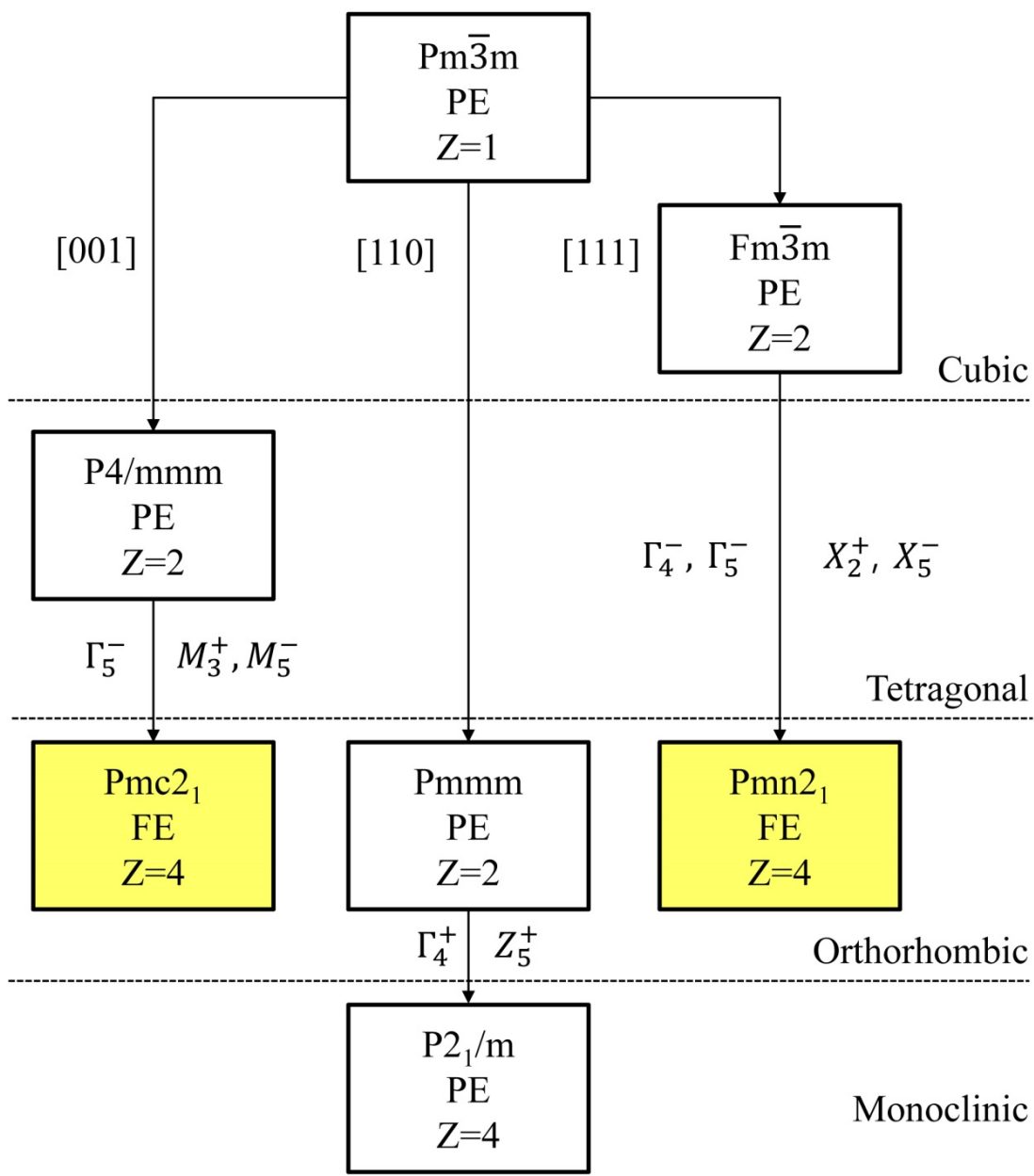

Figure 2 (Color online) Song *et al.*

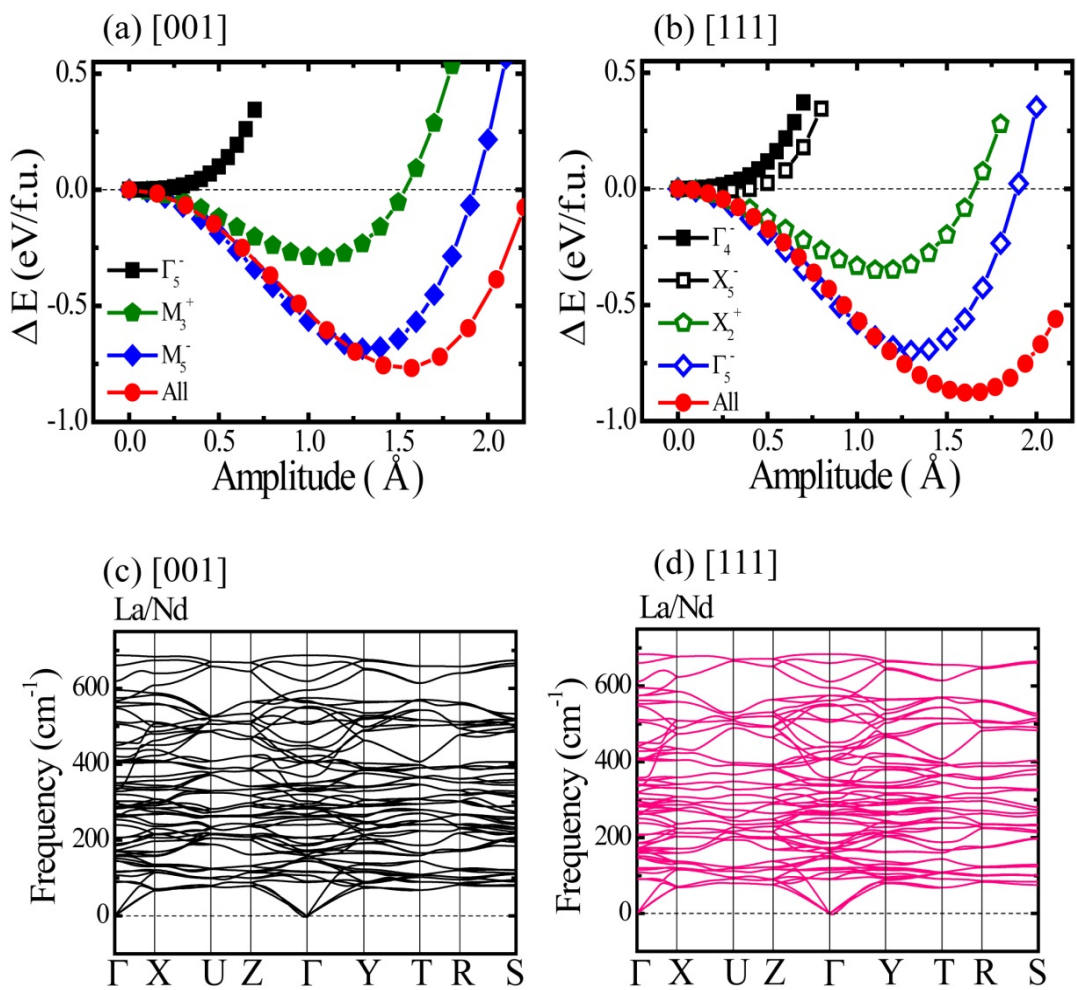

Figure 3 (Color online) Song *et al.*

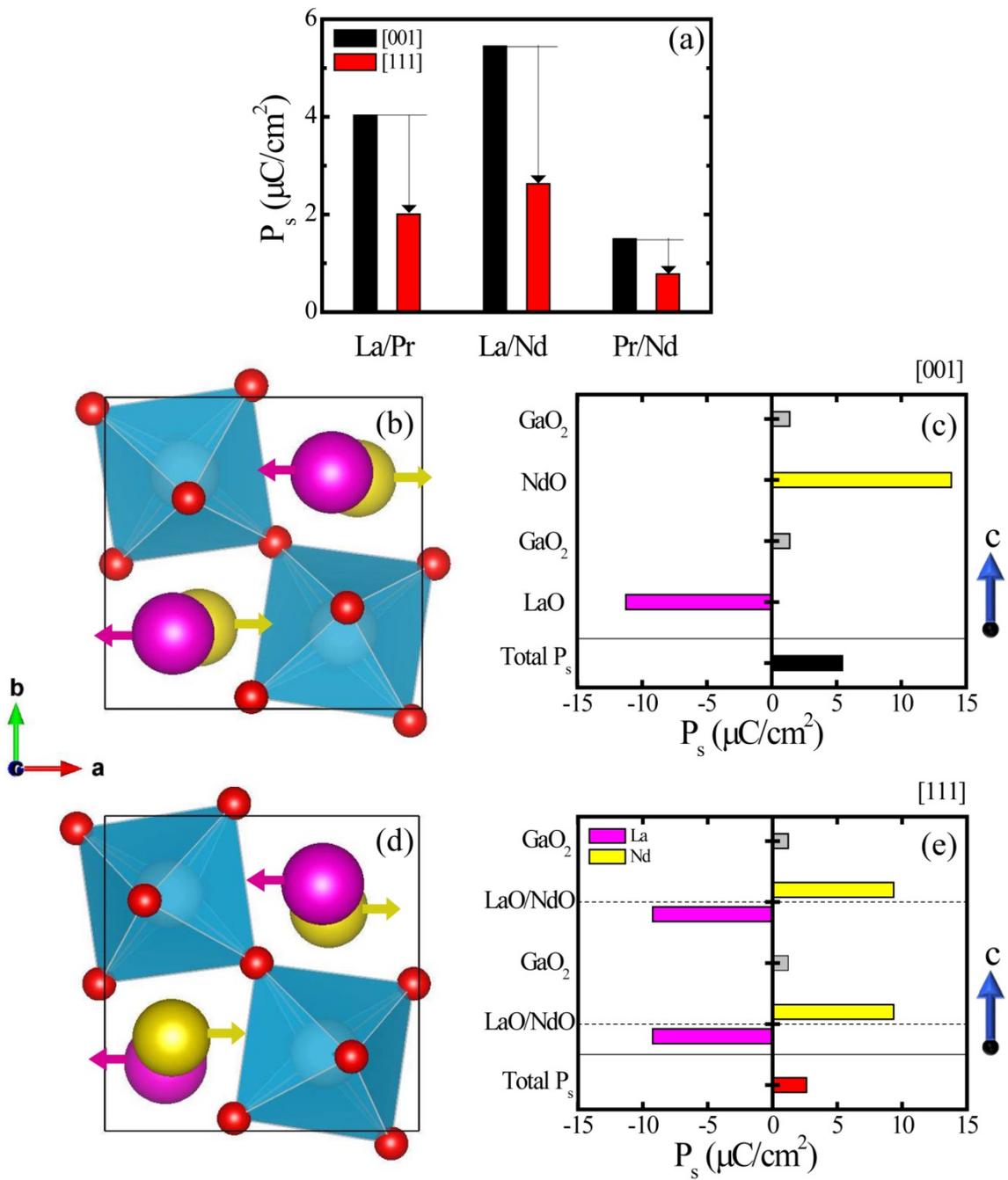

Figure 4 (Color online) Song *et al.*

Table 1. Structural properties of $AGaO_3/A'GaO_3$ (A,A'=La,Pr,Nd) perovskite superlattices with three types cation arrangements: [001] layered ordering, [110] columnar ordering, and [111] rocksalt ordering. $\Delta E_{hl}$ is the energy difference between the high symmetry phase and the low symmetry phase and $\Delta E_{layer}$ is the energy difference between the [001] ordering and the other ordering. Tilting angles of $a^-$ and $c^+$ and polarization values are given for the optimized structure.

| Structure | Ordering | S.G. | $\Delta E_{hl}$(eV/f.u.) | $\Delta E_{layer}$ (meV) | Tilting angle $a^-$ | Tilting angle $c^+$ | P($\mu C/cm^2$) |
|---|---|---|---|---|---|---|---|
| (La,Pr)Ga$_2$O$_6$ | [001] layered | $Pmc2_1$ | -0.683 | 0 | 8.84 | 7.71 | **4.03** |
| (La,Pr)Ga$_2$O$_6$ | [110] columnar | $P2_1/m$ | -0.677 | 17.4 | 8.69 | 7.94 | - |
| (La,Pr)Ga$_2$O$_6$ | [111] rocksalt | $Pmn2_1$ | -0.681 | 13.1 | 8.86 | 7.61 | 2.01 |
| (La,Nd)Ga$_2$O$_6$ | [001] layered | $Pmc2_1$ | -0.804 | 0 | 9.12 | 8.38 | **5.45** |
| (La,Nd)Ga$_2$O$_6$ | [110] columnar | $P2_1/m$ | -0.796 | 26.0 | 8.88 | 8.71 | - |
| (La,Nd)Ga$_2$O$_6$ | [111] rocksalt | $Pmn2_1$ | -0.808 | 12.3 | 9.17 | 8.21 | 2.63 |
| (Pr,Nd)Ga$_2$O$_6$ | [001] layered | $Pmc2_1$ | -0.961 | 0 | 9.50 | 8.87 | **1.50** |
| (Pr,Nd)Ga$_2$O$_6$ | [110] columnar | $P2_1/m$ | -0.953 | 16.0 | 9.38 | 8.65 | - |
| (Pr,Nd)Ga$_2$O$_6$ | [111] rocksalt | $Pmn2_1$ | -0.954 | 13.9 | 9.51 | 8.85 | 0.78 |

Table 2. Mode analysis of the AGaO$_3$/A'GaO$_3$ (A, A'=La, Pr, Nd) superlattices with [001] layered and [111] rocksalt ordering. Note that three modes appear at the orthorhombic transition for the [001] ordering and four modes for the [111] ordering. The irreducible representation of the relevant modes and characteristics displacement type are also given.

| ordering | K-vector | Character | Amplitude in angstrom (mode) | | |
|---|---|---|---|---|---|
| | | | LaPrGa$_2$O$_6$ | LaNdGa$_2$O$_6$ | PrNdGa$_2$O$_6$ |
| [001] layered | (0,0,0) | Ferroelectric | 0.4301 ($\Gamma_5^-$) | 0.4780 ($\Gamma_5^-$) | 0.5488 ($\Gamma_5^-$) |
| | (1/2,1/2,0) | Rotation ($a^0a^0c^+$) | 0.7414 ($M_2^+$) | 0.7982 ($M_3^+$) | 0.8495 ($M_3^+$) |
| | (1/2,1/2,0) | Tilting ($a^-a^-c^0$) | 1.2326 ($M_5^-$) | 1.2704 ($M_5^-$) | 1.3174 ($M_5^-$) |
| | Total | | 1.5014 | 1.5748 | 1.6609 |
| [110] columnar | (0,0,0) | Rotation ($a^0a^0c^+$) | 0.7271 ($\Gamma_4^+$) | 0.7820 ($\Gamma_4^+$) | 0.8443 ($\Gamma_4^+$) |
| | (0,0,1/2) | Tilting ($a^-a^-c^0$) | 1.3085 ($Z_5^+$) | 1.3601 ($Z_5^+$) | 1.4272 ($Z_5^+$) |
| | Total | | 1.4970 | 1.5689 | 1.6583 |
| [111] rocksalt | (0,0,0) | Ferroelectric | 0.1082 ($\Gamma_4^-$) | 0.1297 ($\Gamma_4^-$) | 0.1360 ($\Gamma_4^-$) |
| | (0,1,0) | Rotation ($a^0a^0c^+$) + displacement | 0.4253 ($X_5^-$) | 0.5317 ($X_5^-$) | 0.5470 ($X_5^-$) |
| | (0,1,0) | Rotation ($a^0a^0c^+$) | 0.7293 ($X_2^+$) | 0.8567 ($X_2^+$) | 0.8463 ($X_2^+$) |
| | (0,0,0) | Tilting ($a^-a^-c^0$) | 1.2323 ($\Gamma_5^-$) | 1.3442 ($\Gamma_5^-$) | 1.3115 ($\Gamma_5^-$) |
| | Total | | 1.4979 | 1.6854 | 1.6596 |

**Supplementary information**

Nayoung Song[1], James M. Rondinelli[2], and Bog G. Kim[1]

[1]*Department of Physics, Pusan National University, Pusan, 609-735, South Korea*

[2] *Department of Material Science and Engineering, Northwestern University, 2220 Campus Drive, Evanston, IL 60208-3108, USA*

Some of detailed calculation results are summarized here. Other results can be easily obtainable using the parameters presented in the manuscript. INPUT POSCAR files for optimized structure can be downloaded too.

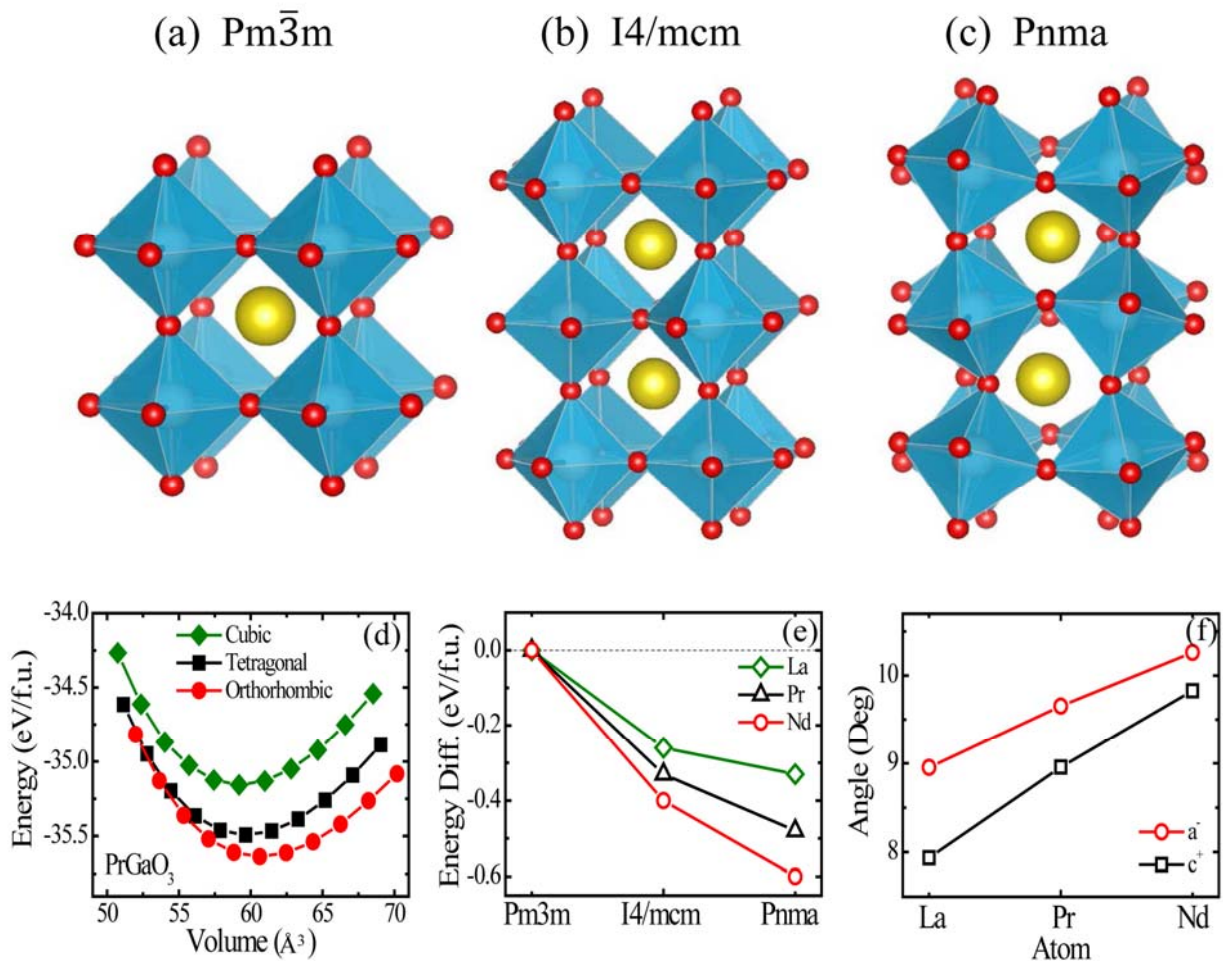

Figure S1. The detailed structure of AGaO$_3$ perovskite with various octahedral tilting, (a) P$m\bar{3}m$ cubic without any tilting, (b) I$4/mcm$ tetragonal with $a^0a^0c^-$ tilting, (c) P$nma$ orthorhombic with $a^-a^-c^+$ tilting. (d) Total energy vs Volume diagram of PrGaO$_3$ with three different symmetries. (e) Total Energy difference of three different phases of LaGaO$_3$, PrGaO$_3$, and NdGaO$_3$. (f) Tilting angels ($a^-$ and $c^+$) in orthorhombic phase of LaGaO$_3$, PrGaO$_3$, and NdGaO$_3$.

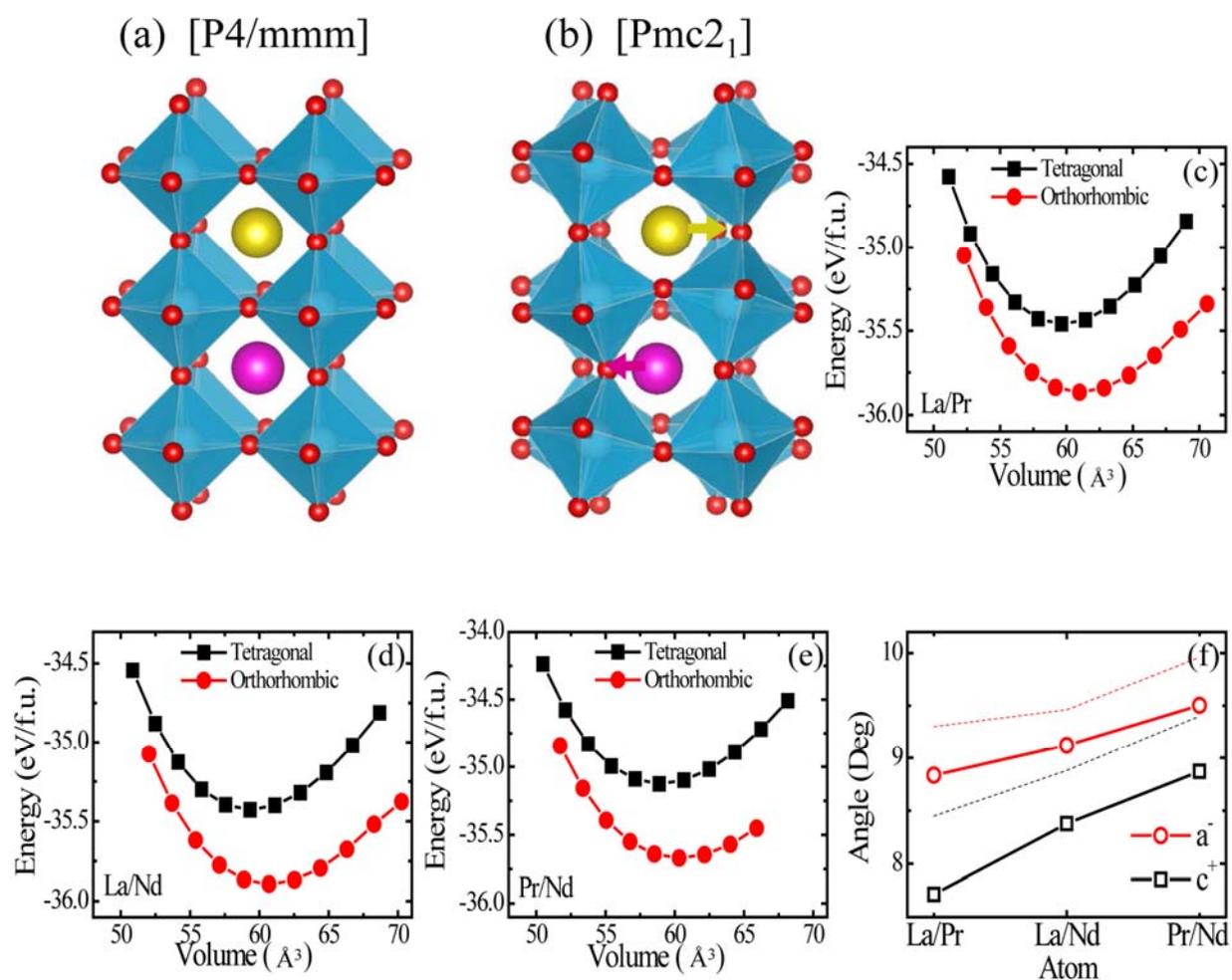

Figure S2. The detailed structure of [001] superlattice AGaO$_3$/A'GaO$_3$ (a) without octahedral tilting (P*4/mmm* tetragonal) and (b) with $a^-a^-c^+$ octahedral tilting (P*mc2$_1$* orthorhombic). Total energy vs volume of tetragonal and orthorhombic phase of (c) LaGaO$_3$/PrGaO$_3$, (d) LaGaO$_3$/NdGaO$_3$, and (e) PrGaO$_3$/NdGaO$_3$ superlattice. (f) Tilting angels ($a^-$ and $c^+$) in the orthorhombic phase of three different superlattices (The dashed lines are average value calculated from the tilting of parent compound).

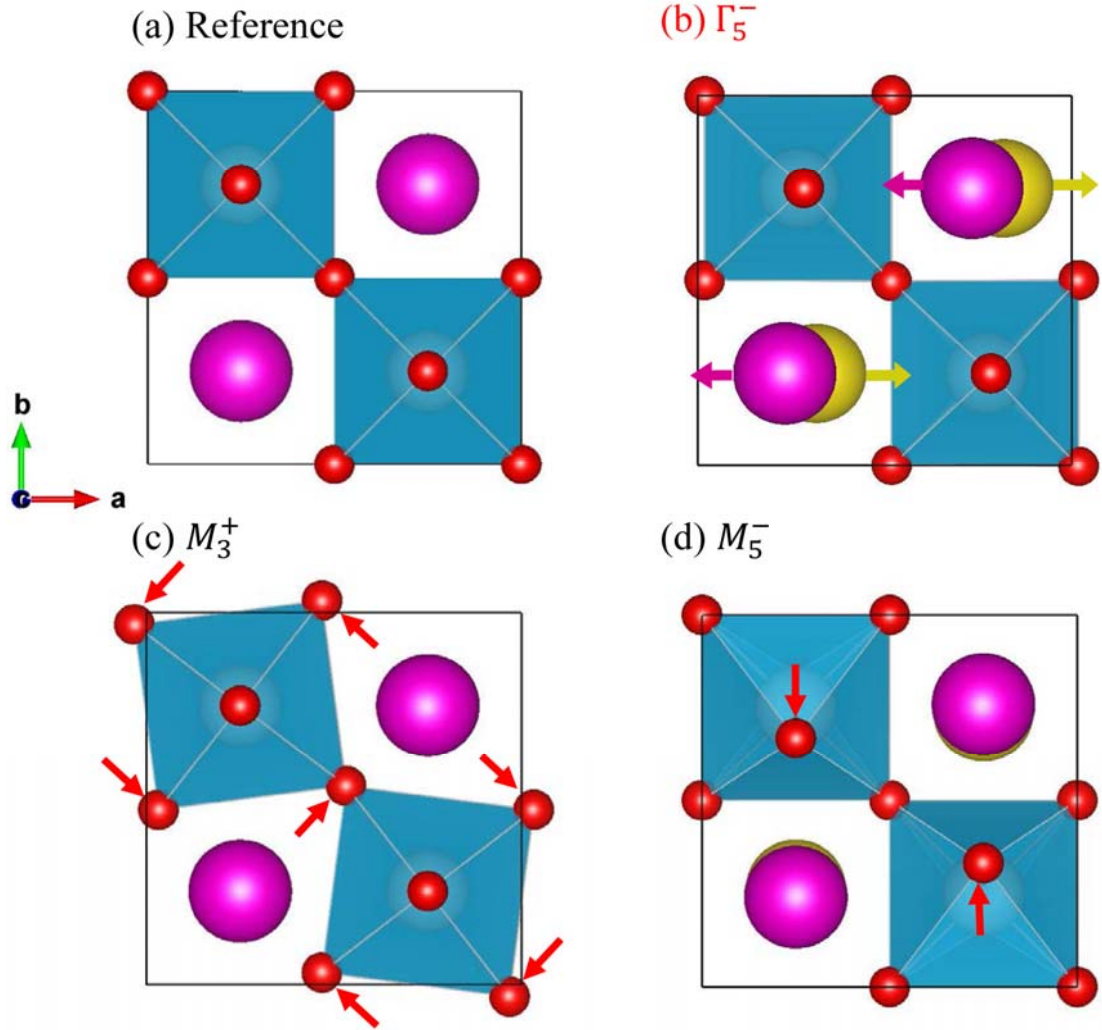

Figure S3. Atomic displacement of irreducible representation of [001] superlattices. (a) The parent compound viewed from c-axis. (b) $\Gamma_5^-$ ferroelectric modes, (c) $M_3^+$ and (d) $M_5^-$ octahedral tilting modes.

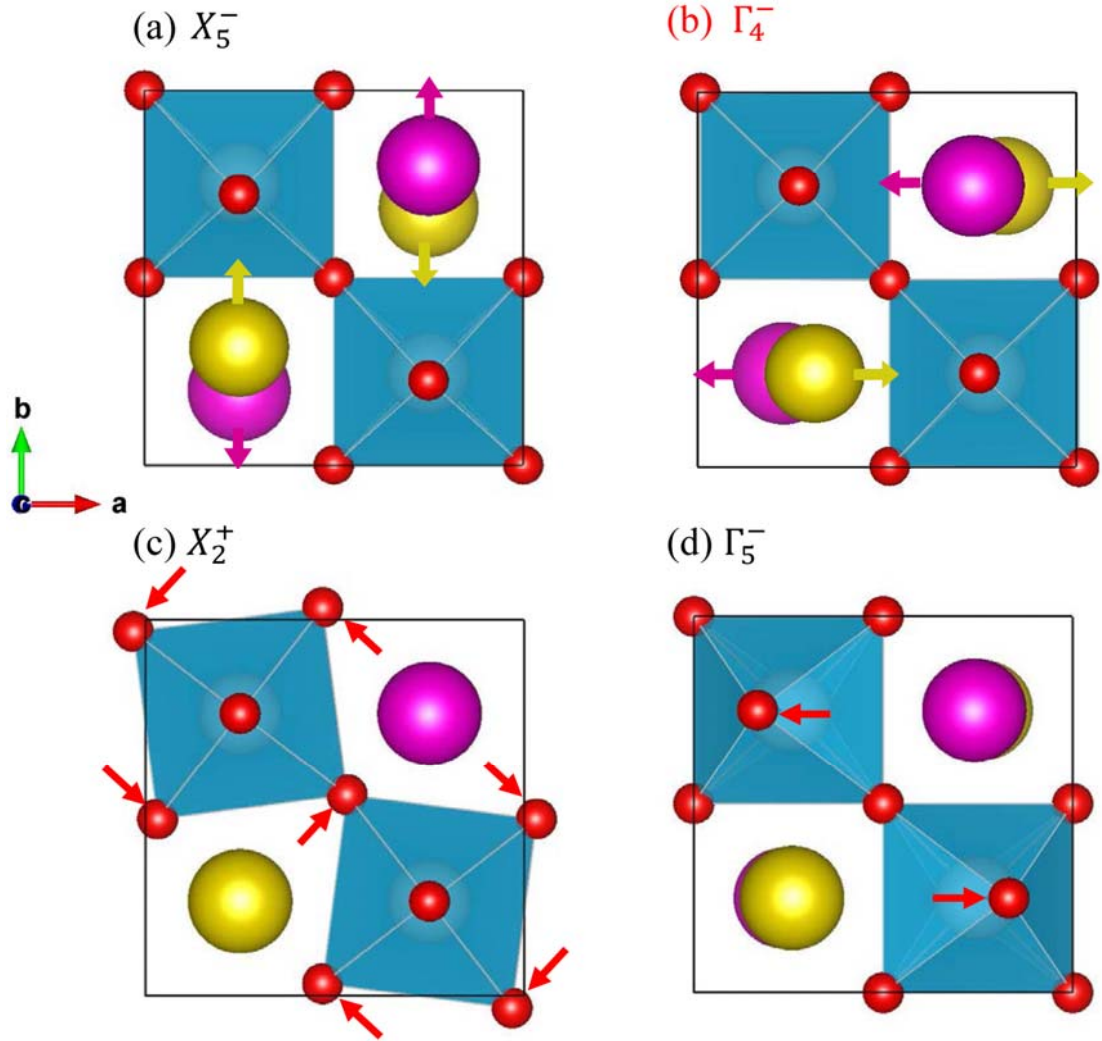

Figure S4. Atomic displacement of irreducible representation of [111] superlattices. (a) $X_5^-$ modes viewed from c-axis. (b) $\Gamma_4^-$ ferroelectric modes, (c) $X_2^+$ and (d) $\Gamma_5^-$ octahedral tilting modes.

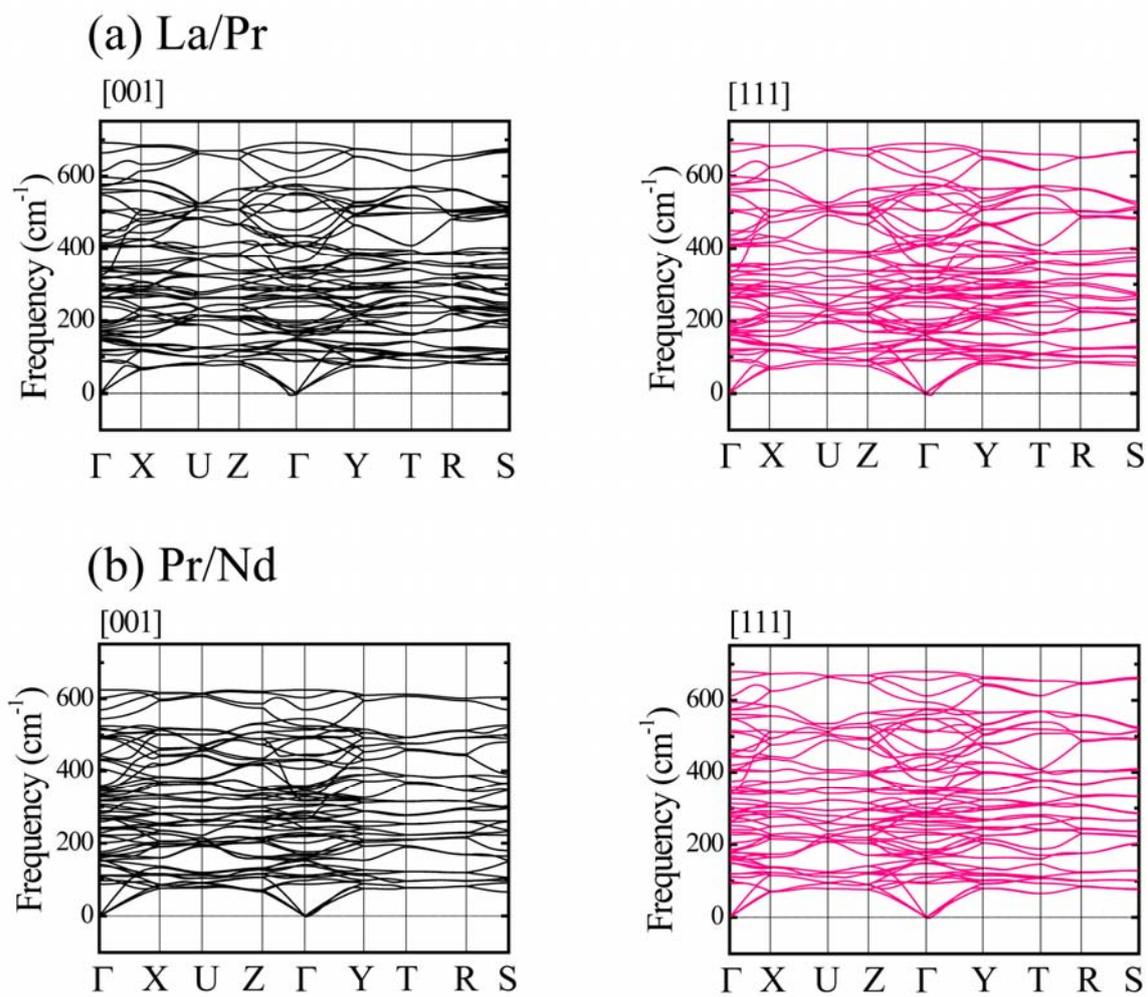

Figure S5. Phonon dispersion curve in the low symmetry phase of (a) La/Pr superlattices (in [001] ordering and [111] ordering) and (b) Pr/Nd superlattices.

| Structure | Tilting angle (deg) | | | ΔE (eV/f.u.) | | |
|---|---|---|---|---|---|---|
| | $a^-$ | $c^+$ | $a^-a^-c^+$ | $Pm\bar{3}m$ | $I4/mcm$ | $Pnma$ |
| LaGaO$_3$ | 8.95 | 7.93 | 12.6 | 0 | -0.26 | -0.33 |
| PrGaO$_3$ | 9.65 | 8.96 | 13.5 | 0 | -0.33 | -0.48 |
| NdGaO$_3$ | 10.26 | 9.83 | 14.4 | 0 | -0.40 | -0.60 |

Table S1. Octahedral tilting angles and other parameters of parent compounds.

Table S2. Wyckoff positions of atoms in the low symmetry phase of AA'Ga$_2$O$_6$ (A, A' = La, Pr, Nd)

| Ordering | Space group | LaPrGa$_2$O$_6$ | | | | |
|---|---|---|---|---|---|---|
| | | Lattice constant | | 5.4762 | 5.5086 | 7.7570 |
| | | SITE | Atom | x | y | z |
| [001] | P*mc*2$_1$ | 2b | La | 0.5000 | 0.7577 | 0.9673 |
| | | 2a | Pr | 0.0000 | 0.7404 | 0.0398 |
| | | 4c | Ga | 0.7512 | 7.7506 | 0.5000 |
| | | 2a | O | 0.0000 | 0.8289 | 0.4836 |
| | | 2b | O | 0.5000 | 0.6760 | 0.5131 |
| | | 4c | O | 0.7934 | 0.4677 | 0.2831 |
| | | 4c | O | 0.7107 | 0.0347 | 0.7142 |
| | | Lattice constant | | 5.4778 | 5.5063 | 7.7552 |
| | | SITE | Atom | x | y | z |
| [110] | P2$_1$/m | 2e | La | 0.4920 | 0.9662 | 0.2500 |
| | | 2e | Pr | 0.0034 | 0.4619 | 0.2500 |
| | | 2a | Ga | 0.0000 | 0.0000 | 0.0000 |
| | | 2d | Ga | 0.5000 | 0.5000 | 0.0000 |
| | | 4f | O | 0.2815 | 0.2175 | 0.0425 |
| | | 4f | O | 0.7842 | 0.2851 | 0.9594 |
| | | 2e | O | 0.9250 | 0.0153 | 0.2500 |
| | | 2e | O | 0.5785 | 0.5140 | 0.2500 |
| | | Lattice constant | | 5.4773 | 5.5071 | 7.7549 |
| | | SITE | Atom | x | y | z |
| [111] | P*mn*2$_1$ | 2a | La | 0.4938 | 0.2180 | 0.0000 |
| | | 2a | Pr | 0.0111 | 0.7102 | 0.0000 |
| | | 4b | Ga | 0.0012 | 0.2481 | 0.7497 |
| | | 4b | O | 0.2817 | 0.4680 | 0.7908 |
| | | 4b | O | 0.7158 | 0.0358 | 0.7078 |
| | | 2a | O | 0.9269 | 0.2656 | 0.0000 |
| | | 2a | O | 0.5802 | 0.7635 | 0.0000 |

| Ordering | Space group | LaNdGa$_2$O$_6$ | | | | |
|---|---|---|---|---|---|---|
| | | Lattice constant | | 5.4568 | 5.5124 | 7.7435 |
| | | SITE | Atom | x | y | z |
| [001] | P$mc2_1$ | 2b | La | 0.7586 | 0.9648 | 0.5000 |
| | | 2a | Nd | 0.7393 | 0.0450 | 0.0000 |
| | | 4c | Ga | 0.7509 | 0.5012 | 0.7523 |
| | | 2a | O | 0.8336 | 0.4802 | 0.0000 |
| | | 2b | O | 0.6757 | 0.5133 | 0.5000 |
| | | 4c | O | 0.4652 | 0.2857 | 0.7962 |
| | | 4c | O | 0.0376 | 0.7115 | 0.7110 |
| | | Lattice constant | | 5.4594 | 5.5113 | 7.7395 |
| | | SITE | Atom | x | y | z |
| [110] | P$2_1/m$ | 2e | La | 0.4913 | 0.9632 | 0.7500 |
| | | 2e | Nd | 0.0106 | 0.4574 | 0.7500 |
| | | 2a | Ga | 0.0000 | 0.0000 | 0.5000 |
| | | 2d | Ga | 0.5000 | 0.5000 | 0.5000 |
| | | 4f | O | 0.2834 | 0.2158 | 0.5430 |
| | | 4f | O | 0.7876 | 0.2885 | 0.4574 |
| | | 2e | O | 0.9243 | 0.0196 | 0.7500 |
| | | 2e | O | 0.5829 | 0.5133 | 0.7500 |
| | | Lattice constant | | 5.4579 | 5.5135 | 7.7389 |
| | | SITE | Atom | x | y | z |
| [111] | P$mn2_1$ | 2a | La | 0.4935 | 0.2162 | 0.0000 |
| | | 2a | Nd | 0.0127 | 0.7045 | 0.0000 |
| | | 4b | Ga | 0.0012 | 0.2473 | 0.7496 |
| | | 4b | O | 0.2829 | 0.4669 | 0.7913 |
| | | 4b | O | 0.7116 | 0.0399 | 0.7058 |
| | | 2a | O | 0.9267 | 0.2684 | 0.0000 |
| | | 2a | O | 0.5851 | 0.7647 | 0.0000 |

| Ordering | Space group | PrNdGa$_2$O$_6$ | | | | |
|---|---|---|---|---|---|---|
| | | Lattice constant | | 5.4285 | 5.5266 | 7.7209 |
| | | SITE | Atom | x | y | z |
| [001] | P*mc*2$_1$ | 2b | Pr | 0.7616 | 0.9543 | 0.5000 |
| | | 2a | Nd | 0.7382 | 0.0483 | 0.0000 |
| | | 4c | Ga | 0.7504 | 0.5064 | 0.7511 |
| | | 2a | O | 0.8343 | 0.4799 | 0.000 |
| | | 2b | O | 0.6701 | 0.5188 | 0.5000 |
| | | 4c | O | 0.4615 | 0.2896 | 0.7958 |
| | | 4c | O | 0.0387 | 0.7104 | 0.7073 |
| [110] | P2$_1$/m | Lattice constant | | 5.4302 | 5.5254 | 7.7200 |
| | | SITE | Atom | x | y | z |
| | | 2e | Pr | 0.4886 | 0.9540 | 0.7500 |
| | | 2e | Nd | 0.0119 | 0.4526 | 0.7500 |
| | | 2a | Ga | 0.0000 | 0.0000 | 0.5000 |
| | | 2d | Ga | 0.5000 | 0.5000 | 0.5000 |
| | | 4f | O | 0.2876 | 0.2115 | 0.5437 |
| | | 4f | O | 0.7891 | 0.2902 | 0.4552 |
| | | 2e | O | 0.9197 | 0.0205 | 0.7500 |
| | | 2e | O | 0.5840 | 0.5162 | 0.7500 |
| [111] | P*mn*2$_1$ | Lattice constant | | 5.4293 | 5.5259 | 7.7195 |
| | | SITE | Atom | x | y | z |
| | | 2a | Pr | 0.4899 | 0.2050 | 0.0000 |
| | | 2a | Nd | 0.0132 | 0.7014 | 0.0000 |
| | | 4b | Ga | 0.0010 | 0.2494 | 0.7499 |
| | | 4b | O | 0.2879 | 0.4620 | 0.7935 |
| | | 4b | O | 0.7110 | 0.0408 | 0.7051 |
| | | 2a | O | 0.9212 | 0.2693 | 0.0000 |
| | | 2a | O | 0.5853 | 0.7675 | 0.0000 |